\documentclass[%
 preprint,
showpacs,
nofootinbib,
 amsmath,amssymb,
 prc,
 superscriptaddress
]{revtex4-1}

\usepackage{graphicx}
\usepackage{dcolumn}
\usepackage{bm}
\usepackage{braket}
\usepackage{caption}
\usepackage{subcaption}
\usepackage{mathrsfs}

\begin{document}


\title{
$\alpha$ Clustering slightly above ${}^{100}$Sn in the Light of the New Experimental Data on the Superallowed $\alpha$ Decay
}


\author{Dong Bai}
 \email{dbai@itp.ac.cn}
\affiliation{School of Physics, Nanjing University, Nanjing 210093, China}%

\author{Zhongzhou Ren}
\email{Corresponding Author: zren@tongji.edu.cn}
\affiliation{School of Physics Science and Engineering, Tongji University, Shanghai 200092, China}%



\date{\today}

\begin{abstract}
The recently observed $\alpha$-decay chain ${}^{108}\text{Xe}\to{}^{104}\text{Te}\to{}^{100}\text{Sn}$ [K.~Auranen \emph{et al.}, Phys.\ Rev.\ Lett.\ {\bf121}, 182501 (2018)] could provide valuable information on the $\alpha$ clustering in even-even nuclei slightly above the major shells $Z=50$ and $N=50$. In this work, this $\alpha$-decay chain is studied theoretically within the framework of the density-dependent cluster model plus the two-potential approach. We calculate the $\alpha$-decay half-lives of ${}^{104}$Te and ${}^{108}$Xe and study in detail their dependence on the $Q$ value and the density-profile parameters of the core nucleus. Various physical properties of ${}^{104}$Te, the heaviest nucleus with a doubly magic self-conjugate core $+$ $\alpha$, are calculated, with two different assumptions on the renormalization factor of the double-folding potential, which could be a useful reference for future experimental studies.
\end{abstract}

\maketitle


\section{Introduction}

$\alpha$ clustering exists in various elements across the nuclide chart. For the light elements, the existence of $\alpha$ clustering has been known for a long time. The relevant studies are revived particularly after the proposal of $\alpha$ condensates in self-conjugate nuclei at the beginning of the twenty-first century \cite{Tohsaki:2001an}. The most famous example of such exotic states is the Hoyle state in $^{12}$C, which is also crucial for the evolution of life on earth. $\alpha$-condensate states are conjectured to exist in heavier self-conjugate nuclei as well. See, e.g., Ref.~\cite{Bai:2018gqt,Katsuragi:2018qaj} and Ref.~\cite{Barbui:2018sqy} for recent theoretical and experimental studies. $\alpha$ clustering also plays an important role in medium-mass, heavy and superheavy nuclei. Valuable information could be obtained by studying the rich $\alpha$-decay and charge-radius data \cite{Delion:1900zz,Delion:2018rrl,Ren:2018xpt,Qi:2018idv,Bai:2018bjl}, which allows the portrait of the landscape of $\alpha$ formation probability and the identification of various (sub)shell effects.

These years, several new $\alpha$ emitters have been observed in the vicinity of the heaviest doubly magic self-conjugate nucleus $^{100}$Sn, such as ${}^{105}$Te, ${}^{106}$Te, ${}^{110}$Xe, and ${}^{114}$Ba \cite{Seweryniak:2006ay,Liddick:2006,Capponi:2016bbw}. The properties of these $\alpha$ emitters have been investigated systematically by one of the authors (ZR) and his collaborator using the density-dependent cluster model (DDCM) \cite{Xu:2006ja}. Good agreements are achieved between theoretical results and experimental data. Other theoretical analyses could be found in, e.g., Ref.~\cite{Mohr:2007,Betan:2012rx,Patial:2016ifz}. Recently, a new $\alpha$-decay chain ${}^{108}\text{Xe}\to{}^{104}\text{Te}\to{}^{100}\text{Sn}$ is observed experimentally by K.~Auranen \emph{et al.}, with two new $\alpha$ emitters $^{108}$Xe [$Q_\alpha=4.6(2)$ MeV, $T_{1/2}=58^{+106}_{-23}$ $\mu$s] and ${}^{104}$Te [$Q_\alpha=5.1(2)$ MeV, $T_{1/2}<18$ ns] being produced by the fusion-evaporation reaction ${}^{54}\text{Fe}({}^{58}\text{Ni},4n){}^{108}\text{Xe}$ \cite{Auranen:2018}. This is the first time to observe an $\alpha$-decay process to a heavy self-conjugate nucleus, and will obviously deepen our understanding of $\alpha$ clustering around the doubly magic numbers and the $N=Z$ line. Noticeably, this $\alpha$-decay chain has already been studied theoretically in Ref.~\cite{Xu:2006ja}. As ${}^{104}$Te and ${}^{108}$Xe have not been discovered yet, in that work their $Q_\alpha$ values are taken from the finite-range droplet model (FRDM) \cite{Moller:1997bz}.

 In the light of the new experimental data, we reanalyze the properties of the $\alpha$-decay chain ${}^{108}\text{Xe}\to{}^{104}\text{Te}\to{}^{100}\text{Sn}$ using a modified DDCM slightly different from Ref.~\cite{Xu:2006ja}, as well as its implication on $\alpha$ clustering in the vicinity of $^{100}$Sn. Especially, we study various physical properties of the new $\alpha$ emitter $^{104}$Te, the heaviest nucleus with a doubly magic self-conjugate core $+$ $\alpha$, which could be a useful reference for future experimental studies. In Section \ref{TF}, we introduce the theoretical framework of our work. In Section \ref{NR}, the numerical results are given. Section \ref{Concl} ends this paper with conclusions and remarks.

\section{Theoretical Framework}
\label{TF}

In the DDCM, the parent nucleus is treated as a binary system made of the $\alpha$ particle and the core nucleus. The central potential is given by:
\begin{align}
V(r)=V_{\alpha\text{-core}}(r)+\frac{\hbar^2}{2\mu}\frac{L(L+1)}{r^2},
\end{align}
with the first term $V_{\alpha\text{-core}}(r)$ being the effective potential between the $\alpha$ particle and the core nucleus, and the second term being the centrifugal potential. $\mu$ is the two-body reduced mass. In literature, various $\alpha$-core effective potentials have been proposed, such as the Cosh potential \cite{Buck:1992zz}, the Woods-Saxon (WS) potential \cite{Denisov:2005ax,Ni:2009vzd}, the $\text{WS}+\text{WS}^3$ potential \cite{Buck:1995zza}, the double-folding potential \cite{Xu:2005ukj}, the Woods-Saxon-Gaussian potential \cite{Bai:2018hbe}, etc. In this work, we take the double-folding potential to describe the $\alpha$-core effective interaction, which is given explicitly by
\begin{align}
&V_{\alpha\text{-core}}(\mathbf{r})=V_\text{N}(\mathbf{r})+V_\text{C}(\mathbf{r}),\\
&V_\text{N}(\mathbf{r})=\lambda\int\mathrm{d}\mathbf{r}_\alpha\mathrm{d}\mathbf{r}_c\,\rho_\alpha(\mathbf{r}_\alpha)\rho_c(\mathbf{r}_c)v_n(Q_\alpha, \mathbf{s}\equiv\mathbf{r}+\mathbf{r}_c-\mathbf{r}_\alpha),\label{VN}\\
&V_\text{C}(\mathbf{r})=\int\mathrm{d}\mathbf{r}_\alpha\mathrm{d}\mathbf{r}_c\,\tilde{\rho}_\alpha(\mathbf{r}_\alpha)\tilde{\rho}_c(\mathbf{r}_c)v_c(\mathbf{s}\equiv\mathbf{r}+\mathbf{r}_c-\mathbf{r}_\alpha),\\
&v_n(Q_\alpha,\mathbf{s})=7999\frac{\exp(-4s)}{4s}-2134\frac{\exp(-2.5s)}{2.5s}+276(0.005Q_\alpha/A_\alpha-1)\delta(\mathbf{s}),\\
&v_c(s)=\frac{e^2}{s}.
\end{align}
Here, $\rho_\alpha(\mathbf{r}_\alpha)$ and $\rho_c(\mathbf{r}_c)$ ($\tilde{\rho}_\alpha(\mathbf{r}_\alpha)$ and $\tilde{\rho}_c(\mathbf{r}_c)$) are the nucleon (proton) density profiles of the $\alpha$ particle and the core nucleus. $v_n$ is the M3Y nucleon-nucleon interaction derived from the Reid soft-core potential with the last term being the exchange component, and has been widely used in the theoretical studies of heavy-ion scatterings \cite{Satchler:1979}. $v_c$ is the proton-proton Coulomb interaction. $\lambda\sim0.5-0.9$ \cite{Xu:2006fq} is the renormalization factor introduced phenomenologically to achieve a better agreement between theoretical results and experimental data. 

Given the $\alpha$-core potential $V(r)$, the energy spectrum of the target nucleus could be obtained by solving the two-body Schr\"odinger equation in the quasibound-state approximation. The $\alpha$-decay half-life is, on the other hand, could be estimated by different methods such as the WKB approximation \cite{Delion:1900zz}, the two-potential approach and its modification \cite{Gurvitz:1987,Gurvitz:1988,Gurvitz:2004}, the GAMOW code \cite{Vertse:1982}, etc. Here, we adopt the two-potential approach,
\begin{align}
T_{1/2}=\frac{\hbar \ln2}{\Gamma_\alpha},\qquad \Gamma_\alpha=P_\alpha\frac{4\hbar^2\tilde{k}^2}{\mu k}\left|\phi_L(R)\chi_L(kR)\right|^2,
\label{TG}
\end{align}
where $P_\alpha$ is the $\alpha$ formation probability, $k=\sqrt{2\mu Q_\alpha}/\hbar$, and $\tilde{k}=\sqrt{2\mu (V(R)-Q_\alpha)}/\hbar$. $\chi_{L}(kR)$ is the regular eigenfunction of the outer Hamiltonian and can be approximated by the standard regular Coulomb function in practical calculations. $\phi_{L}(R)$ is the radial wave function. Here, $R$ is the separation radius of the two-potential method and is chosen to satisfy $V'(R)=0$. The error caused by this choice would be tiny for our purpose. See Ref.~\cite{Gurvitz:2004} for more discussions on the choice of the separation radius. For simplicity, the $\alpha$ formation probability is taken to be $P_\alpha=1$, from which more realistic choices could be easily implemented. To measure the agreement between the theoretical results and experimental values, a hindrance factor is defined as follows:
\begin{align}
\mathcal{HF}=\frac{T_{1/2}^\text{th}}{T_{1/2}^\text{exp}}.
\label{SF}
\end{align}
In literature, the hindrance factor could also be interpreted as a probe of the $\alpha$ formation probability, thus giving a convenient measure of the strength of $\alpha$ clustering of the target nucleus \cite{Brown:1992rg}.

We will study the electromagnetic transitions among different states in the ground-state band of ${}^{104}\text{Te}$. The $\gamma$-decay width for the quadrupole transition from the initial state with the angular momentum $L$ to the final state with angular momentum $L-2$ could be obtained by
\begin{align}
&\Gamma_\gamma(\text{E2};L\to L-2)=\frac{12\pi e^2}{225}\left(\frac{E_\gamma}{\hbar c}\right)^5B(\text{E2};L\to L-2),\\
&B(\text{E2};L\to L-2)=\frac{15\beta_2^2}{8\pi}\frac{L(L-1)}{(2L+1)(2L-1)}\left|\int_0^\infty \mathrm{d}r\,\phi_{L-2}(r)^*r^2\phi_{L}(r)\right|^2,\\
&\beta_2=e\frac{Z_cA_\alpha^2+Z_\alpha A_c^2}{(A_c+A_\alpha)^2}.
\end{align}
Here, we have ignored the contributions from the internal conversion to $\Gamma_\gamma$ for simplicity. Then, the branching ratio for the $\alpha$ decay is given by
\begin{align}
b_\alpha=\frac{\Gamma_\alpha}{\Gamma_\alpha+\Gamma_\gamma}.
\end{align}
The relative sizes between electromagnetic transitions and $\alpha$ decays may play a crucial role in preparing ${}^{104}$Te in its ground state experimentally. If $\Gamma_\gamma$ is too small, ${}^{104}$Te produced by the fusion-evaporation reaction in excited states might decay directly to the ${}^{100}$Sn ground state through $\alpha$ decay, which would make it difficult to study the ground-state $\alpha$ decay of ${}^{104}$Te.

\section{Numerical Results}
\label{NR}

To calculate the double-folding potential between the $\alpha$ particle and the core nucleus ${}^{100}$Sn and ${}^{104}$Te, we take the following density profiles
\begin{align}
&\rho_\alpha(r)=0.422875\exp(-0.7024r^2),\label{GF}\\
&\rho_c(r)=\frac{\rho_0}{1+\exp\left(\frac{r-c}{a}\right)}.\label{FF}
\end{align} 
The density profile for the $\alpha$ particle is taken to be the Gaussian form, which is consistent with the experimental data from the electron scattering \cite{Satchler:1979}. With the density profile Eq.~\eqref{GF}, the root-mean square (RMS) charge radius of the $\alpha$ particle is found to be \cite{Satchler:1979}
\begin{align}
r_\text{ch}=\sqrt{\braket{r^2}_\text{p}+0.76-0.11(N/Z)}=1.67\text{ fm}.\label{CR}
\end{align}
On the right-hand side of Eq.~\eqref{CR}, the first term $\braket{r^2}_p=\int\mathrm{d}\mathbf{r}\,r^2\rho_\alpha(r)/\int\mathrm{d}\mathbf{r}\rho_\alpha(r)$ is the point radius of the $\alpha$ particle, the second term 0.76 $\text{fm}^2$ corresponds to the charge radius squared of the proton, and the third term $-0.11$ $\text{fm}^2$ corresponds to the charge radius squared of the neutron. The obtained theoretical value on the charge radius of the $\alpha$ particle is consistent with the experimental one given by Ref.~\cite{Angeli:2013}. We assume the core nuclei ${}^{100}$Sn and ${}^{104}$Te to be approximately spherical, which is supported by theoretical studies based on the FRDM, and take their density profile to be the standard Fermi form \cite{Bohr:1998}. The normalization $\rho_0$ is determined by the mass number $A_c$ of the core nucleus, and the other constants are first taken as \cite{Bohr:1998}
\begin{align}
\text{$c=1.07 A_c^{1/3}$ and $a=0.54$ fm.}
\label{BV}
\end{align}
Later on, we will vary these parameters and study their impacts on the results. The corresponding RMS charge radius is found to be $r_\text{ch}=4.41$ fm for ${}^{100}$Sn and $r_\text{ch}=4.46$ fm for ${}^{104}$Te. It is interesting to compare these two values with the theoretical values given by the empirical formula \cite{Angeli:2013}
\begin{align}
r^\text{em}_\text{ch}=\left(0.9071+\frac{1.105}{A_c^{2/3}}-\frac{0.548}{A_c^{4/3}}\right)A_c^{1/3},
\label{EF}
\end{align}
which are found to be $r^\text{em}_\text{ch}=4.44$ fm for ${}^{100}$Sn and $r^\text{em}_\text{ch}=4.50$ fm for ${}^{104}$Te. With the density profiles in Eq.~\eqref{GF} and \eqref{FF}, the double-folding potential could be calculated for the $\alpha$ particle and the core nucleus. More details on calculating the double-folding potential could be found in, e.g., the Appendix of Ref.~\cite{Bai:2018hbe}.

The Pauli principle, which plays a fundamental role in nuclear many-body physics, is simulated by the so-called Wildermuth condition \cite{Wildermuth:1977}
\begin{align}
G=2N+L=\sum_{i=1}^4(2n_i+l_i),
\end{align} 
where $G$ is the global quantum number, $N$ is the number of nodes in the relative wave function, and $L$ is the relative orbital angular momentum of the $\alpha$-core binary system. $n_i$ and $l_i$ are the corresponding quantum numbers of the nucleons in the $\alpha$ particle. For ${}^{104}$Te and ${}^{108}$Xe, the four valence nucleons in the $\alpha$ particle all occupy the orbit $0g_{7/2}$, with $n_i=0$ and $l_i=4$. As a result, we take $G=16$.

To determine the renormalization factor $\lambda$, we require the $\alpha$-core effective potential $V_{\alpha\text{-core}}(r)$ to be able to reproduce the experimentally measured $Q$ values for $L=0$, i.e., $Q_\alpha=4.6(2)$ MeV for ${}^{108}$Xe and $Q_\alpha=5.1(2)$ MeV for ${}^{104}$Te. Concerning calculating the energy spectrum from the double-folding potential, there are, at least, two possible choices. The first is to do the calculation with the renormalization factor being constant. Sometimes, this choice would result in the inverted band for heavy elements such as ${}^{212}$Po, which apparently contradicts the experimental data. The other choice is to make the renormalization factor depend on the angular momentum $L$ \cite{Ohkubo:1995},
\begin{align}
\lambda(L)=\lambda(0)-g \times L,\qquad g\sim10^{-3}.
\label{Lg}
\end{align}
 This choice would not only fix the problem of the inverted band but also help achieve a better phenomenological agreement. We will take both choices in studying the ground-state band of ${}^{104}$Te.
 
 The numerical results are given as follows. We first take the benchmark value Eq.~\eqref{BV} for the density profile of the core nucleus. As shown by Table \ref{BR}, the theoretical results are consistent with the experimental values. The upper (lower) limit of the theoretical result is obtained by taking the $Q$ value to be its lower (upper) limit. For the center values, we have $T^\text{th}_{1/2}({}^{108}\text{Xe})=28\times10^3$ ns for $Q_\alpha({}^{108}\text{Xe})=4.6$ MeV and $T^\text{th}_{1/2}({}^{104}\text{Te})=32$ ns for $Q_\alpha({}^{104}\text{Te})=5.1$ MeV.  
 
 
\begin{table}
\caption{Benchmark results for the $\alpha$-decay chain ${}^{108}\text{Xe}\to{}^{104}\text{Te}\to{}^{100}\text{Sn}$ with the density-profile parameters taken to be Eq.~\eqref{BV}. In calculating the hindrance factors, we take $T_{1/2}^\text{exp}({}^{108}\text{Xe})=58\times10^3$ ns, the center value of the current experimental value, and $T_{1/2}^\text{exp}({}^{104}\text{Te})=18$ ns, which is the upper bound of the current experimental measurement.
}
\label{BR}
\begin{center}
\begin{tabular}{ccccccccccccc}
\hline
\hline
\hspace{5mm}$\text{Nucleus}$\hspace{5mm} & \hspace{5mm}{$Q_\alpha$ (MeV)}\hspace{5mm}  & \hspace{5mm}{$T_{1/2}^\text{exp}$ (ns)}\hspace{5mm} & \hspace{5mm}{$T_{1/2}^\text{th}$ (ns)}\hspace{5mm} & \hspace{5mm}{$\mathcal{HF}$}\hspace{5mm}
\\[0.5ex]  
\hline
${}^{108}\text{Xe}\to{}^{104}\text{Te}$ & 4.6(2) & $58^{+106}_{-23}\times10^{3}$ & $(4.1-213)\times10^3$ & $0.07-4$
\\[0.5ex]
${}^{104}\text{Te}\to{}^{100}\text{Sn}$ & 5.1(2) & $<18$ & $7-166$ & $0.4-9$ 
\\
\hline
\hline
\end{tabular}
\end{center}
\end{table}

As a benchmark, we also calculate the physical properties of the rotational band of ${}^{104}$Te using the double-folding potential with the renormalization factor being constant. The results are summarized in Table \ref{RB}. Unlike ${}^{212}$Po, for ${}^{104}$Te we do not encounter the inverted-band problem. In the same table, the $\alpha$-decay width and $\gamma$-decay width for each excited state are given as well. It is found that, with the renormalization factor being constant, the $\gamma$-decay width of the $2^+$ state is found to be a bit smaller than the corresponding $\alpha$-decay width, which might be problematic for producing ${}^{104}$Te in its ground states experimentally.

 \begin{table}
\caption{Benchmark results for the rotational band of ${}^{104}$Te with the renormalization factor $\lambda=0.64417$ being constant with respect to the angular momentum $L$. Here, we take the $Q$ value to be $Q_\alpha({}^{104}\text{Te})=5.1$ MeV and the density-profile parameters Eq.~\eqref{BV}.
}
\label{RB}
\begin{center}
\begin{tabular}{ccccccccccccc}
\hline
\hline
\hspace{5mm}$J^\pi$\hspace{5mm} & \hspace{5mm}{$Q_\alpha$ (MeV)}\hspace{5mm} & \hspace{5mm}{$E_x$ (MeV)}\hspace{5mm} & \hspace{5mm}{$\Gamma_{\alpha}$ (MeV)}\hspace{5mm} & \hspace{5mm}{$\Gamma_\gamma$ (MeV)}\hspace{5mm} & \hspace{5mm}{$b_\alpha(\%)$}\hspace{5mm}
\\[0.5ex]  
\hline
$0^+$ & $5.10$ & $0$ & $1.44\times10^{-14}$ & $-$ & $-$ \\[0.5ex]
$2^+$ & $5.22$ & $0.12$ & $1.59\times10^{-14}$ & $6.31\times10^{-15}$ & 72 \\[0.5ex]
$4^+$ & $5.49$ & $0.39$ & $1.76\times10^{-14}$ & $5.23\times10^{-13}$ & 3 \\[0.5ex]
$6^+$ & $5.91$ & $0.81$ & $1.56\times10^{-14}$ & $4.24\times10^{-12}$ & 0.4 \\[0.5ex]
$8^+$ & $6.46$ & $1.36$ & $8.83\times10^{-15}$ & $1.45\times10^{-11}$ & 0.06 \\[0.5ex]
$10^+$ & $7.12$ & $2.02$ & $2.69\times10^{-15}$ & $3.09\times10^{-11}$ & 0.009 \\[0.5ex]
$12^+$ & $7.88$ & $2.78$ & $3.91\times10^{-16}$ & $4.77\times10^{-11}$ & 0.0008 \\[0.5ex]
$14^+$ & $8.75$ & $3.65$ & $2.45\times10^{-17}$ & $5.63\times10^{-11}$ & 0.00004\\[0.5ex]
$16^+$ & $9.71$ & $4.61$ & $5.18\times10^{-19}$ & $4.60\times10^{-11}$ & $10^{-6}$\\[0.5ex]
\hline
\hline
\end{tabular}
\end{center}
\end{table}

Next, we study the impacts of the $Q$-value uncertainties and varying the density-profile parameters of the core nucleus ${}^{100}$Sn. We first consider the impacts of the $Q$-value uncertainties. Fig.~(1a) (Fig.~(1c)) shows how the $\alpha$-decay half-life of ${}^{104}$Te (${}^{108}\text{Xe}$) calculated by the DDCM varies with respect to the $Q$ value within its error bar of 0.2 MeV. It is found that, as the $Q$ value increases, the $\alpha$-decay half-lives decrease correspondingly. Furthermore, as can be seen in Fig.~(1a), the upper limit of the $\alpha$-decay half-life $T_{1/2}^\text{exp}<18$ ns seems to favor the $Q$ value of ${}^{104}$Te located in the upper part of its experimental error bar. In Fig.~(1b) (Fig.~(1d)), the relation between the $\alpha$-decay half-life of ${}^{104}$Te (${}^{108}$Xe) and the parameter $\lambda$ in Eq.~\eqref{VN} is shown. Following Ref.~\cite{Ni:2013dfa}, we also vary the density-profile parameters $r_0$ and $a$ for the core ${}^{100}$Sn, and find that the $\alpha$-decay half-life of ${}^{104}$Te and the RMS of ${}^{100}$Sn could also be changed correspondingly. The relevant results are shown in Fig.~\ref{VDP}. Compared with the diffuseness parameter $a$, the $\alpha$-decay half-life is more sensitive to the $r_0$ parameter. Also, we find that, when the RMS radius of ${}^{100}$Sn increases (decreases), the $\alpha$-decay half-life decreases (increases). The blue triangles in Fig.~\ref{VDP} correspond to the theoretical values with the RMS radius of ${}^{100}$Sn taken to be $r^\text{em}_\text{ch}=4.44$ fm, which might be helpful for experimentalists.

\begin{figure}
\centering
\begin{subfigure}[b]{\textwidth}
\centering
\minipage{0.48\textwidth}
  \includegraphics[width=\linewidth]{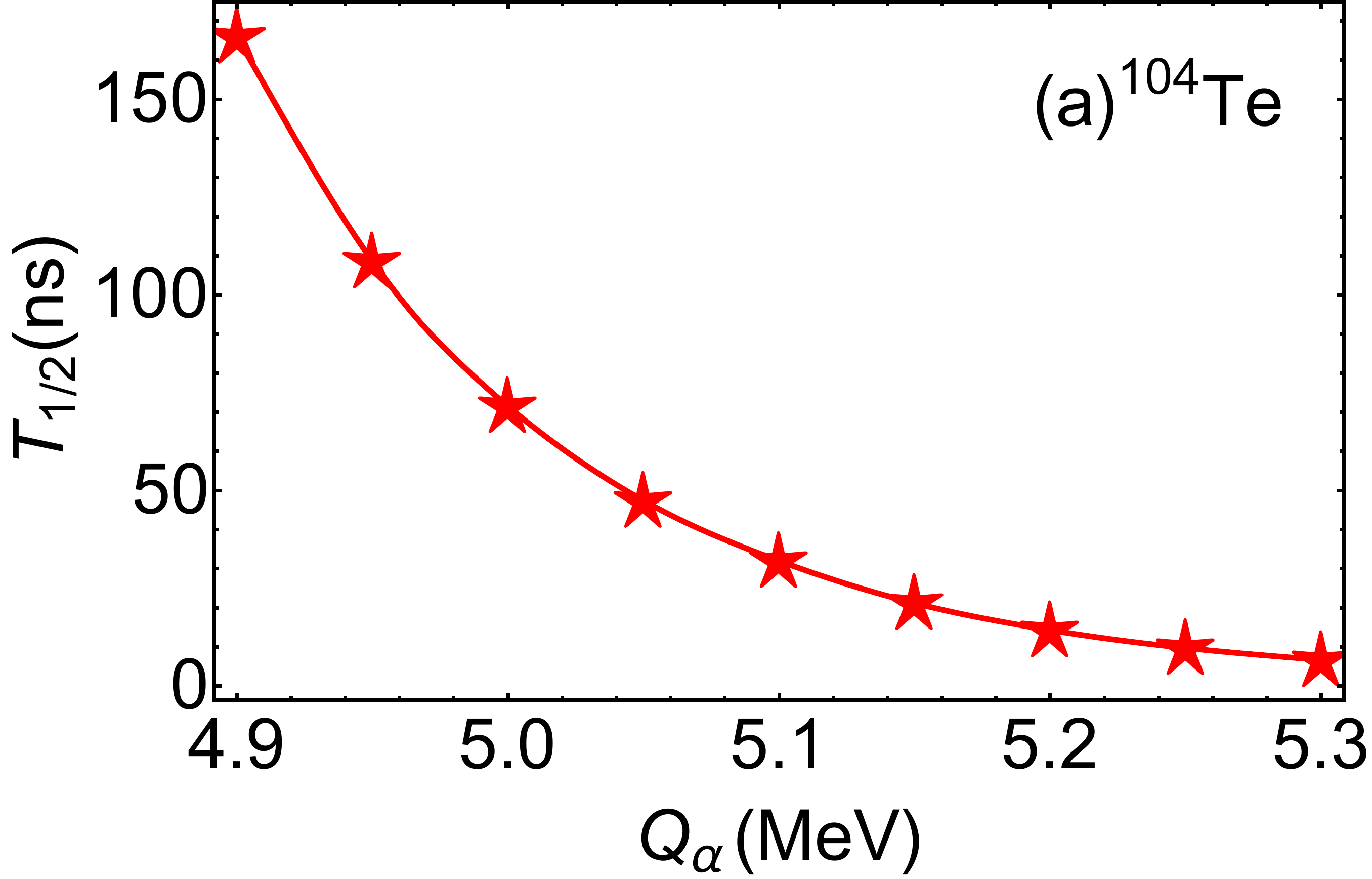}
\endminipage\hfill
\minipage{0.48\textwidth}
  \includegraphics[width=\linewidth]{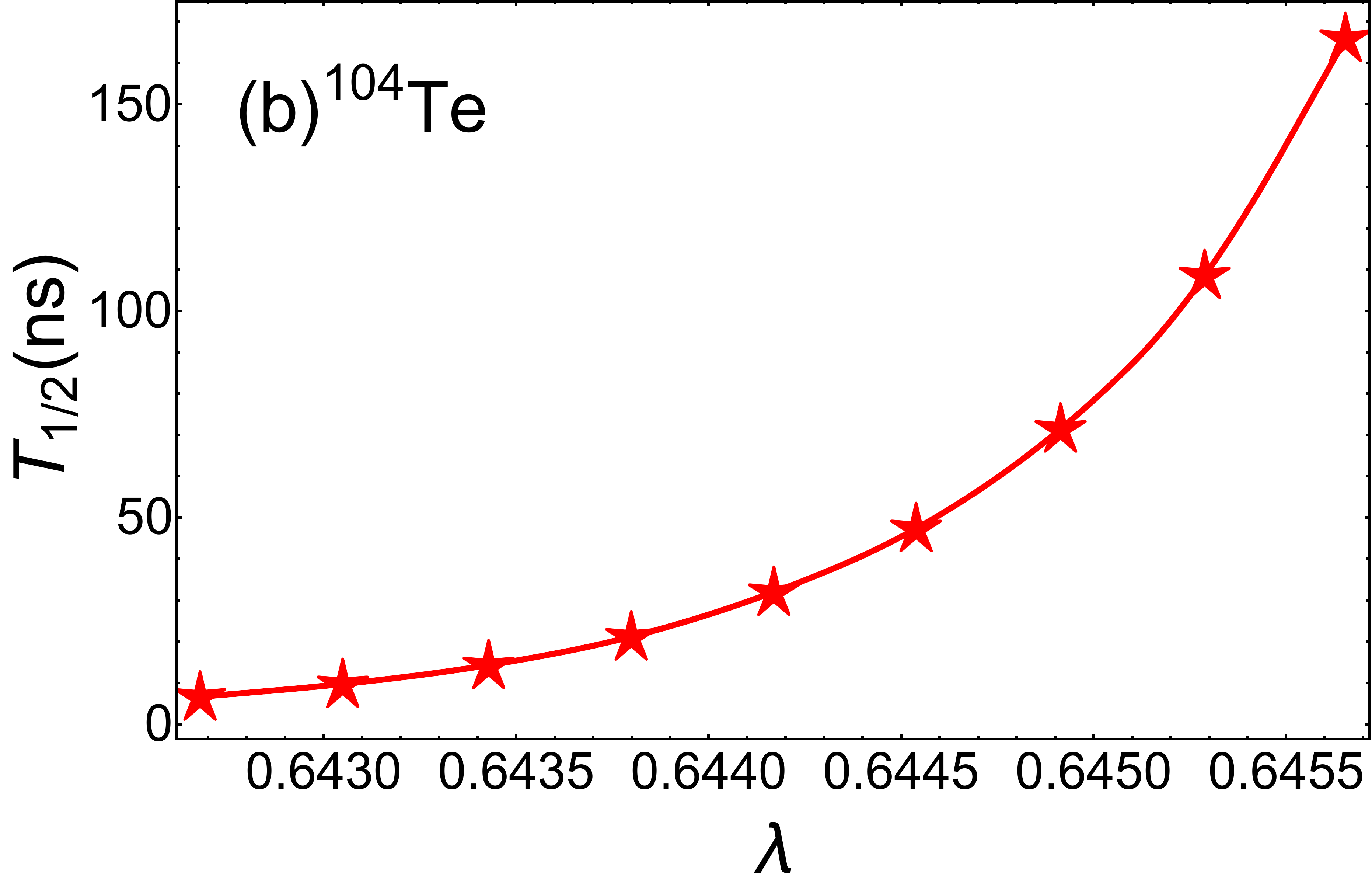}
\endminipage
\end{subfigure}

\begin{subfigure}[b]{\textwidth}
\centering
\minipage{0.48\textwidth}
  \includegraphics[width=\linewidth]{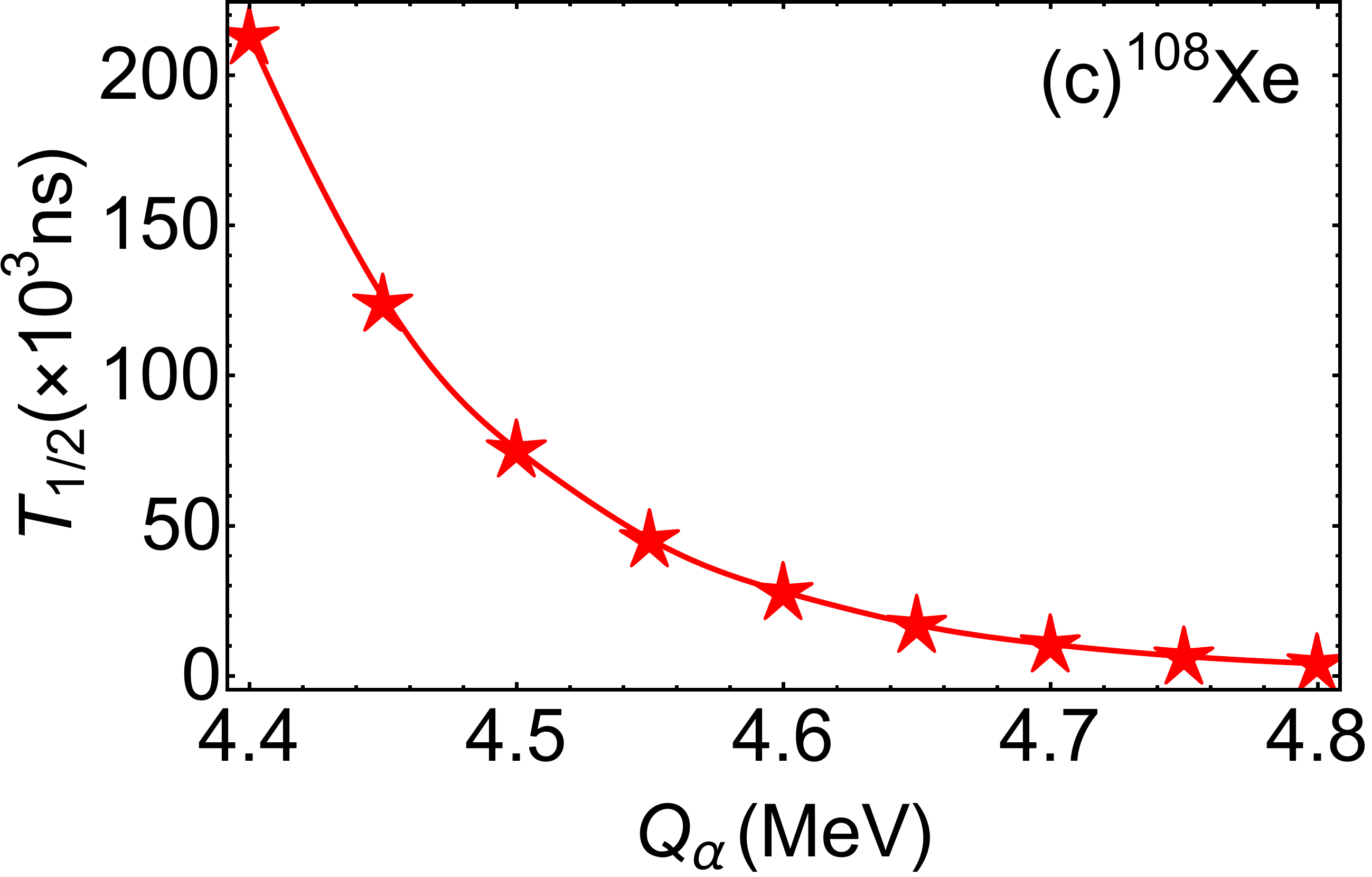}
\endminipage\hfill
\minipage{0.48\textwidth}
  \includegraphics[width=\linewidth]{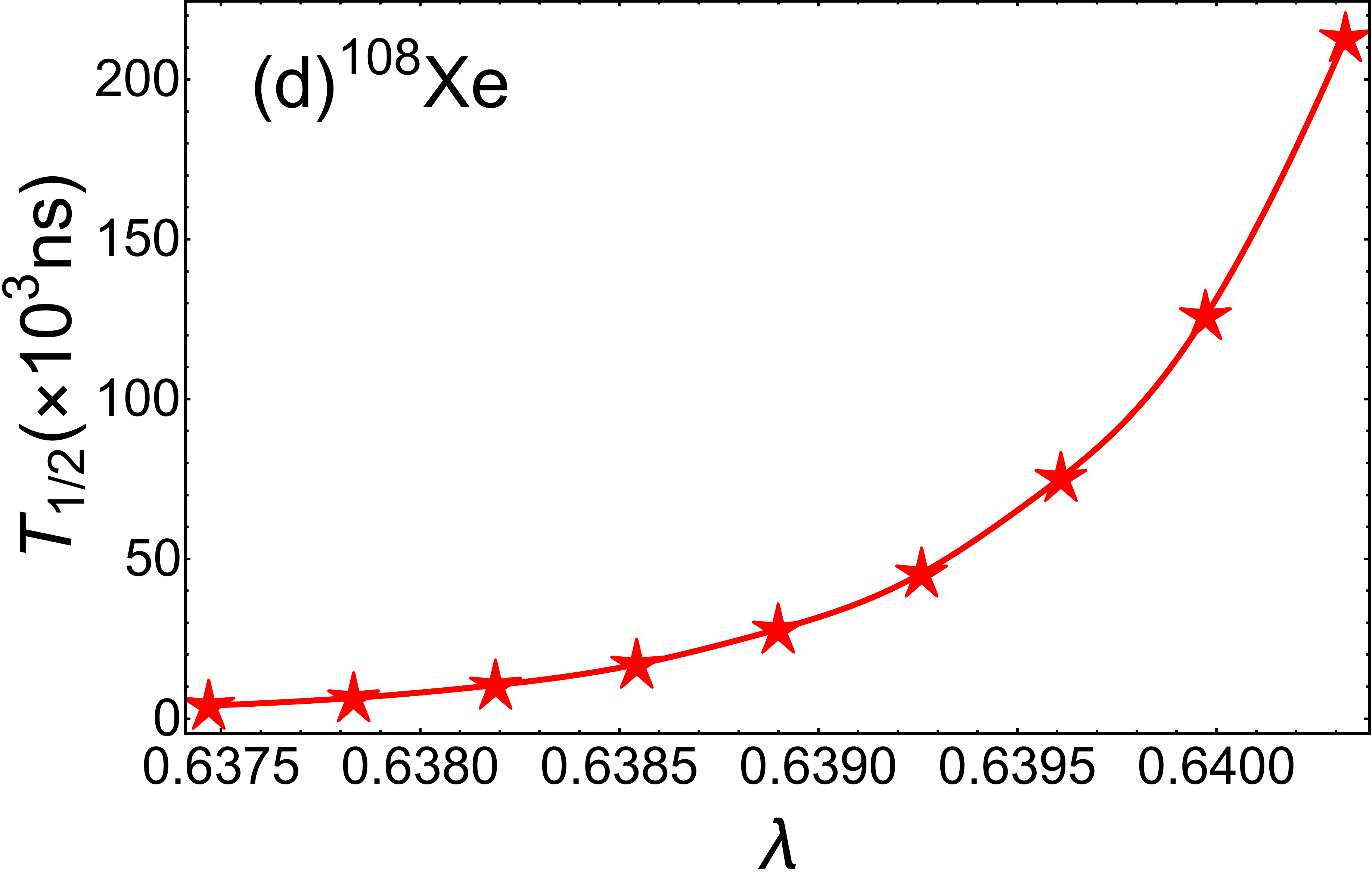}
\endminipage
\end{subfigure}

\caption{(a) The $\alpha$-decay half-life of ${}^{104}$ Te by the DDCM versus the $Q$ value within its error bar of 0.2 MeV. (b) The $\alpha$-decay half-life of ${}^{104}$Te by the DDCM versus the parameter $\lambda$ in Eq.~\eqref{VN}. (c) The same as Fig.~(1a) except for ${}^{108}$Xe. (d) The same as Fig.~(1b) except for ${}^{108}$Xe.}
\label{VQa}
\end{figure}

\begin{figure}
\centering

\begin{subfigure}[b]{\textwidth}
\centering
\minipage{0.48\textwidth}
  \includegraphics[width=\linewidth]{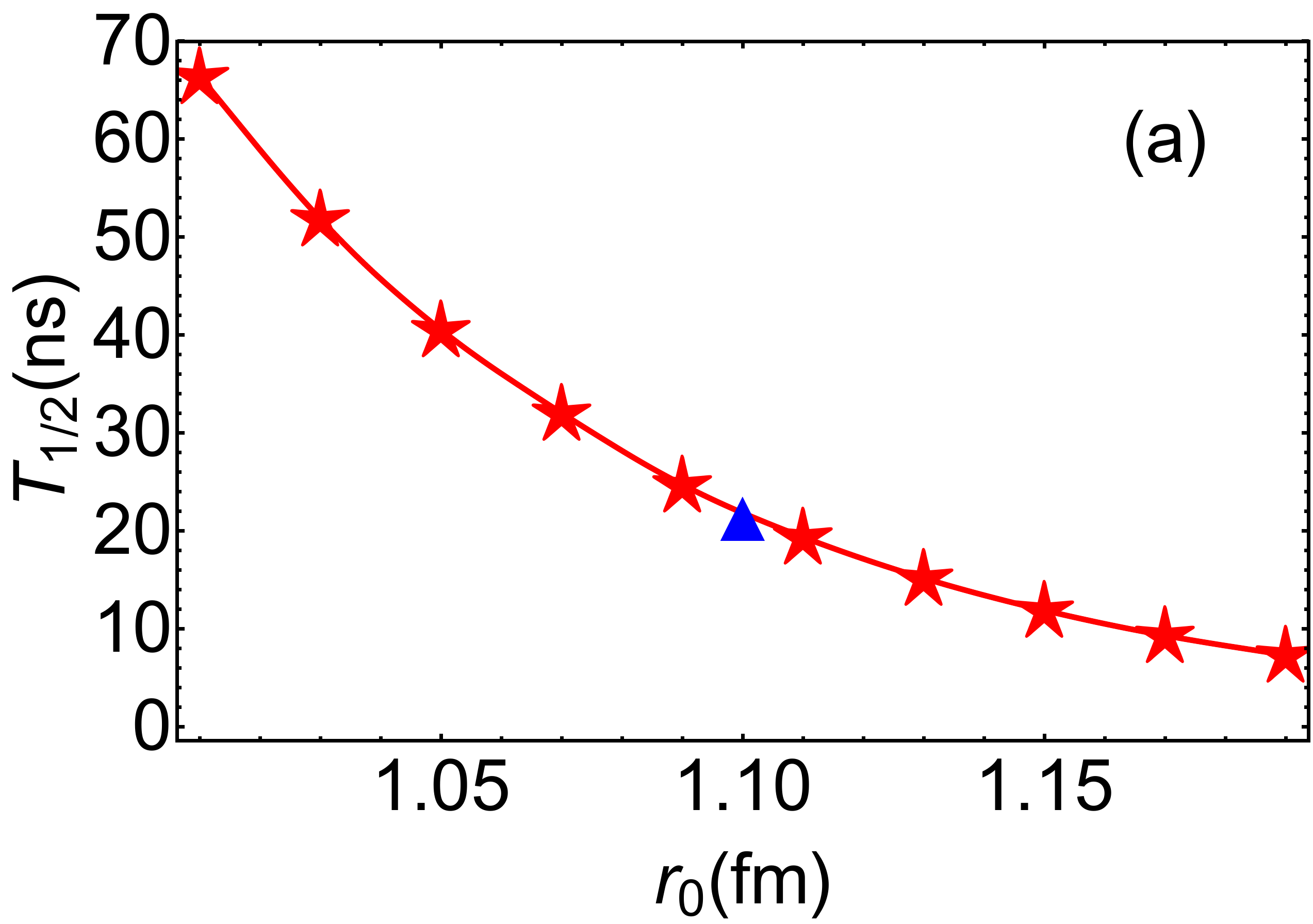}
\endminipage\hfill
\minipage{0.48\textwidth}
  \includegraphics[width=\linewidth]{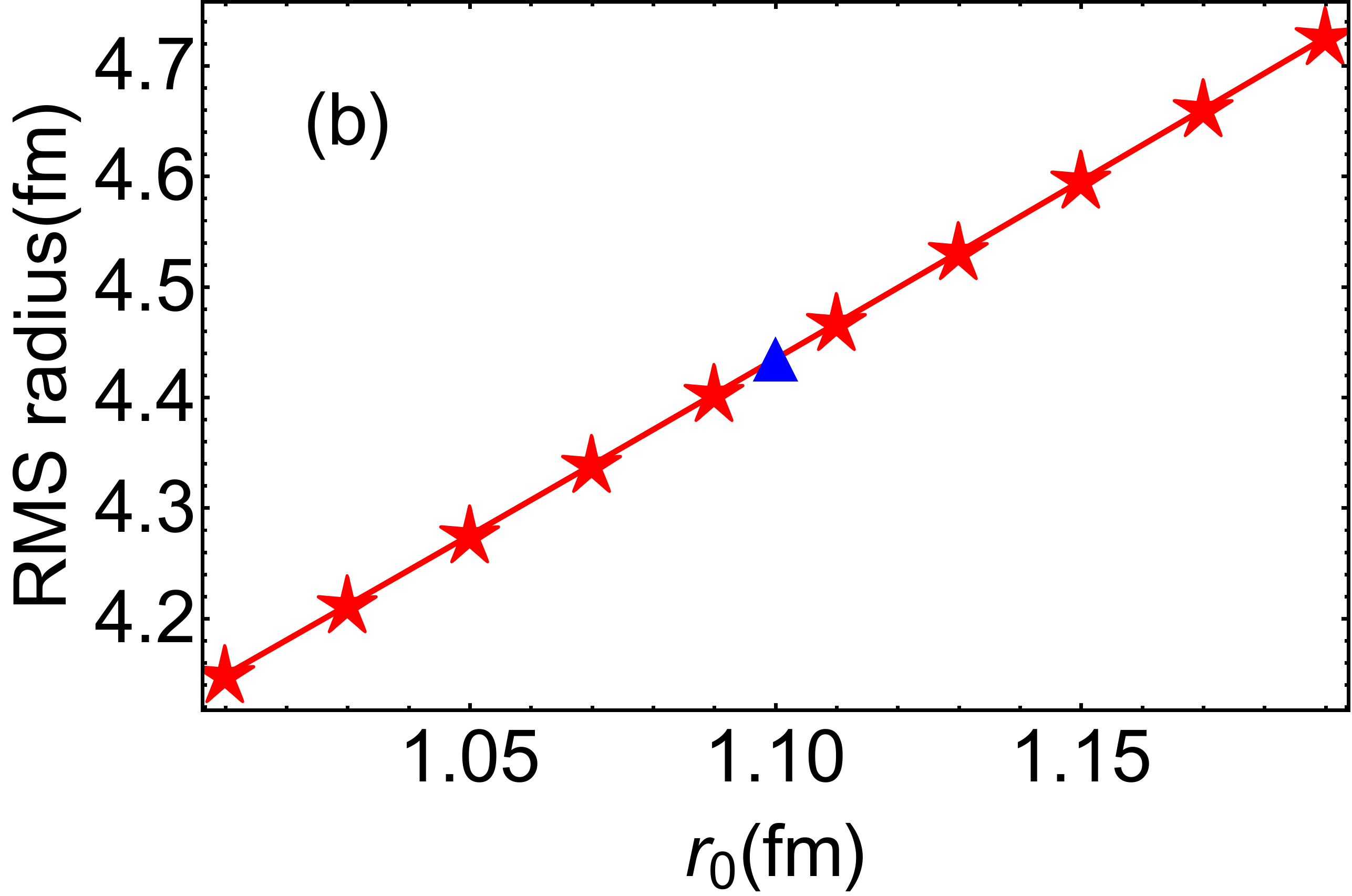}
\endminipage
\end{subfigure}

\begin{subfigure}[b]{\textwidth}
\centering
\minipage{0.48\textwidth}
  \includegraphics[width=\linewidth]{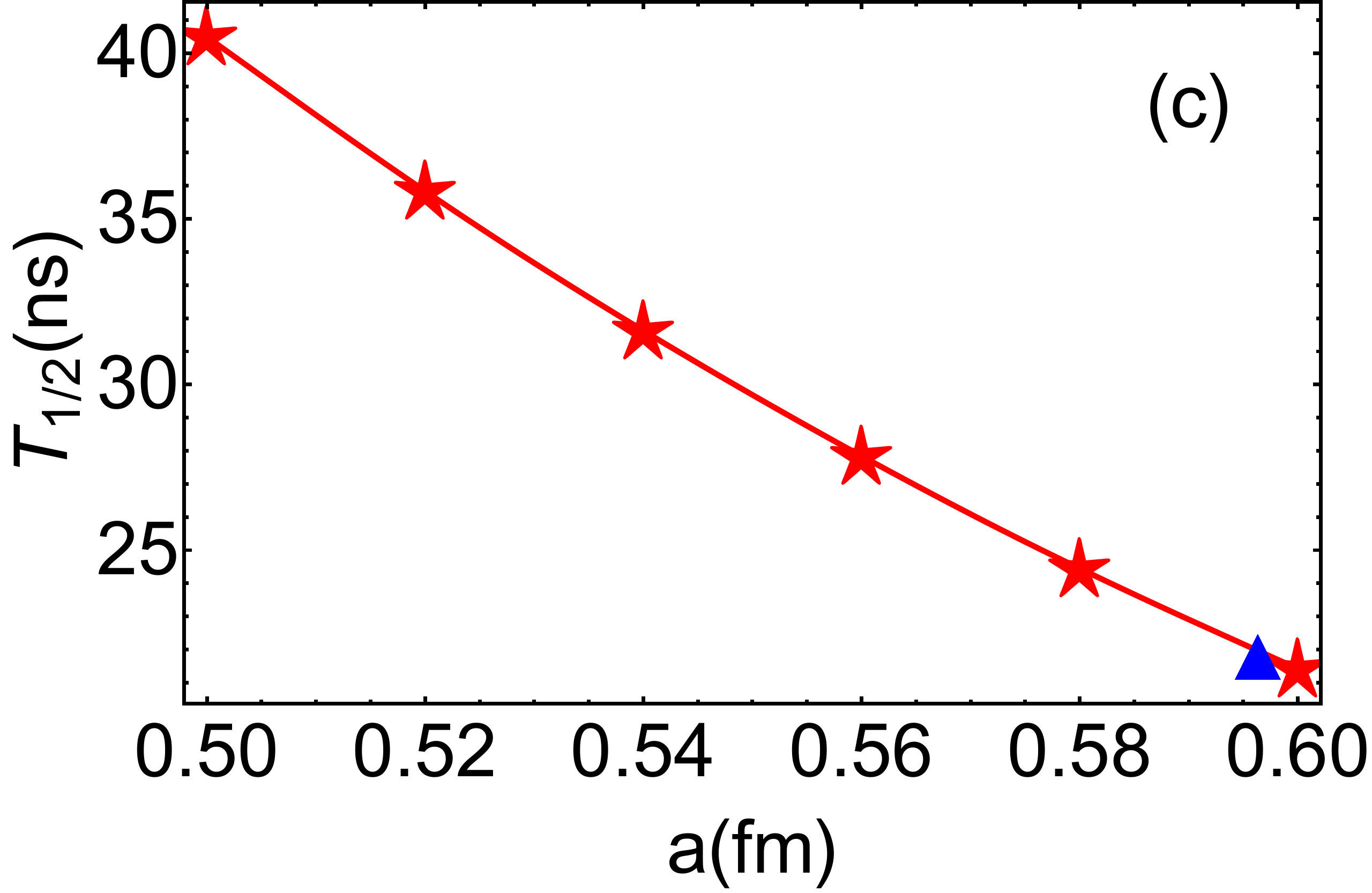}
\endminipage\hfill
\minipage{0.50\textwidth}
  \includegraphics[width=\linewidth]{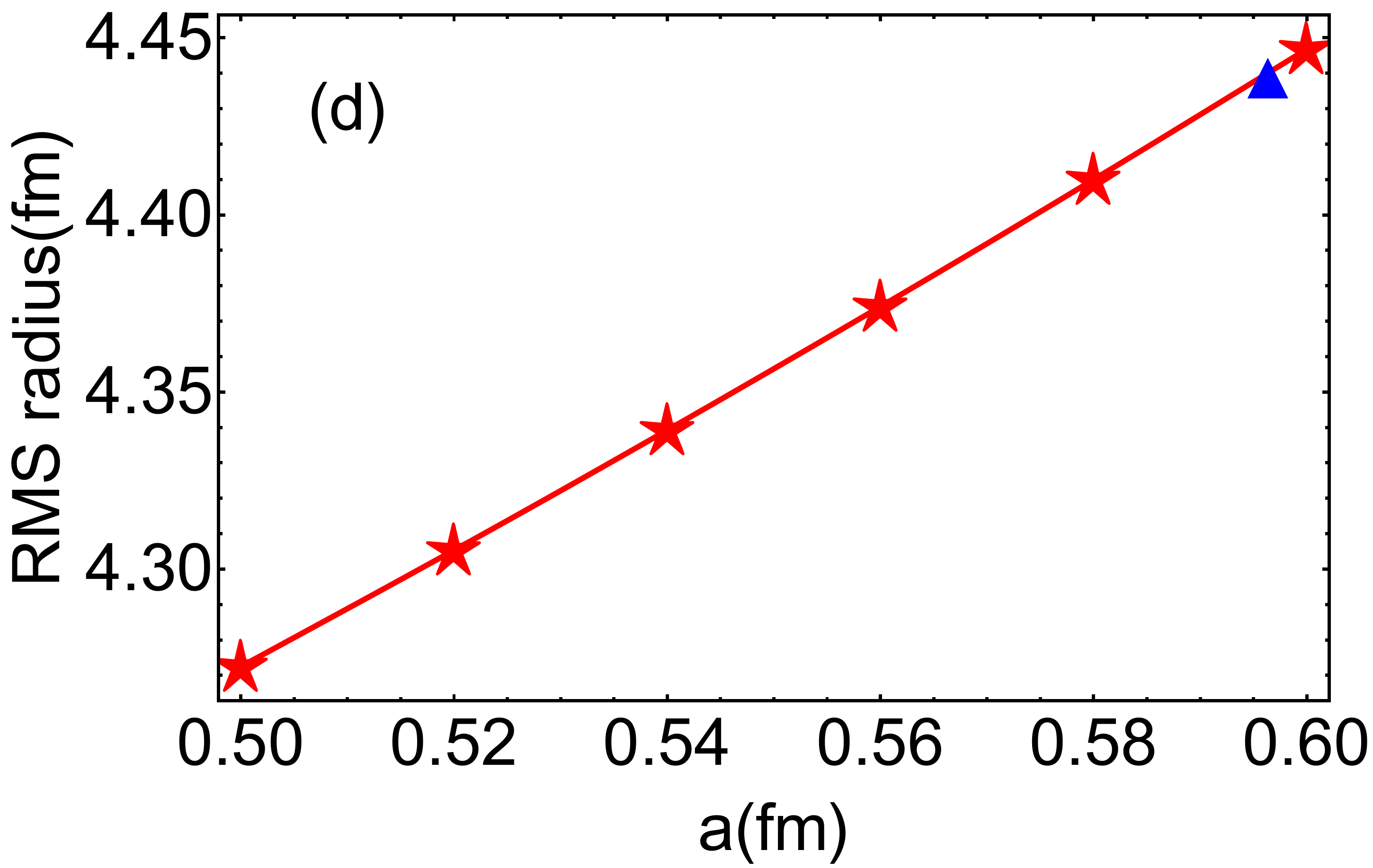}
\endminipage
\end{subfigure}

\caption{Impacts of varying the density-profile parameters on the theoretical results for ${}^{104}$Te. The red stars and curves in Fig.~(2a) and (2b) show the impacts of varying the parameter $r_0$ of the ${}^{100}$Sn density profile on the $\alpha$-decay half-life and the RMS radius of ${}^{100}$Sn, while the blue triangle is the theoretical value corresponding to the empirical RMS charge radius $r^\text{em}_\text{ch}=4.44$ fm given by Eq.~\eqref{EF}. The calculation is done with the fixed diffuseness parameter $a=0.54$ fm. The red stars and curves in Fig.~(2c) and (2d) show the impacts of varying the diffuseness parameter $a$ of the ${}^{100}$Sn density profile on the $\alpha$-decay half-life and the RMS radius of ${}^{100}$Sn, while the blue triangle is the theoretical value corresponding to the empirical RMS charge radius $r^\text{em}_\text{ch}=4.44$ fm given by Eq.~\eqref{EF}. The calculation is done with the fix $r_0$ parameter $r_0=1.07$ fm. }
\label{VDP}
\end{figure}

At last, we consider the impacts of adopting the $L$-dependent renormalization factor for the low-lying excited states of ${}^{104}$Te. We mainly pay attention to the $2^+$ and $4^+$ states and the renormalization factor is given by Eq.~\eqref{Lg}. From Table \ref{RBVL}, one can see that, by introducing the $L$-dependent renormalization factor, the energy levels of the ${2}^{+}$ and $4^+$ states increase compared with the results in Table \ref{RB}. Meanwhile, as the parameter $g$ increases, the $\alpha$-decay branching ratio of the $2^+$ state decreases quickly and the $\gamma$-decay mode starts to dominate, which could be helpful for producing the ${}^{104}$Te in its ground state experimentally.

 \begin{table}
\caption{Energy levels and $\alpha$-decay branching ratios of the $2^+$ and $4^+$ states obtained by varying the renormalization factor of the double-folding potential according to Eq.~\eqref{Lg}.}
\label{RBVL}
\begin{center}
\begin{tabular}{ccccccccccccc}
\hline
\hline
\hspace{5mm}$g (\times10^{-3})$\hspace{5mm} & \hspace{5mm}$J^\pi$\hspace{5mm} & \hspace{5mm}{$E_x$ (MeV)}\hspace{5mm} & \hspace{5mm}{$b_\alpha(\%)$}\hspace{5mm}
\\[0.5ex]  
\hline
0 & $2^+$ &  $0.12$ & $72$ \\[0.5ex]
   & $4^+$ & $0.39$ & $3$ \\[0.5ex]
\hline
1 & $2^+$ &  $0.25$ & $15$ \\[0.5ex]
   & $4^+$ & $0.65$ & $3$ \\[0.5ex]
\hline
2 & $2^+$ &  $0.37$ & $5$ \\[0.5ex]
   & $4^+$ & $0.91$ & $4$ \\[0.5ex]
\hline
3 & $2^+$ &  $0.49$ & $3$ \\[0.5ex]
   & $4^+$ & $1.16$ & $5$ \\[0.5ex]
\hline
4 & $2^+$ &  $0.62$ & $2$ \\[0.5ex]
   & $4^+$ & $1.41$ & $9$ \\[0.5ex]
\hline
5 & $2^+$ &  $0.75$ & $2$ \\[0.5ex]
   & $4^+$ & $1.66$ & $15$ \\[0.5ex]
\hline
\hline
\end{tabular}
\end{center}
\end{table}

\section{Conclusions}
\label{Concl}

In this work, we study in detail the physical properties of two new $\alpha$ emitters ${}^{104}$Te and ${}^{108}\text{Xe}$ in the light of the new experimental data on their $\alpha$ decays. The theoretical investigation is carried out in the framework of the DDCM, with the $\alpha$-core interaction given by the double-folding potential and the $\alpha$-decay half-lives estimated by the two-potential approach. The calculated $\alpha$-decay half-lives are consistent with the new experimental data. Meanwhile, we study the energy spectrum and decay properties of ${}^{104}$Te, which is the heaviest nucleus with a doubly magic self-conjugate nucleus plus the $\alpha$ cluster. We also study the dependence of the theoretical results on the $Q$ value and density-profile parameters of the core nucleus, which could help motivate future experimental studies on these properties of $\alpha$ emitters around ${}^{100}$Sn, which, if available, would clearly help deepen our understanding of $\alpha$ clustering. 

\begin{acknowledgments} 
We would like to thank Peter Mohr for helpful communications, especially for his comments on the $Q$-value uncertainties. This work is supported by the National Key R\&D Program of China (Contract No.~2018YFA0404403, 2016YFE0129300), by the National Natural Science Foundation of China (Grant No.~11535004, 11761161001, 11375086, 11120101005, 11175085 and 11235001), and by the Science and Technology Development Fund of Macau under Grant No.~008/2017/AFJ. DB is also supported by Project funded by China Postdoctoral Science Foundation (Grant No.~2018M640470).
\end{acknowledgments}



\end{document}